# THE OPPORTUNISTIC TRANSMISSON OF WIRELESS WORMS BETWEEN MOBILE DEVICES

C. J. Rhodes[*1] and M. Nekovee[2]

[1]*Institute for Mathematical Sciences, Imperial College London, 53 Prince's Gate, Exhibition Road, South Kensington, London, SW7 2PG, United Kingdom.*

[2]*BT Research, Polaris 134, Adastral Park, Martlesham, Suffolk, IP5 3RE, United Kingdom.*
*and*
*Centre for Computational Science, University College London, 20 Gordon Street, London, WC1H 0AJ, United Kingdom.*

* author for correspondence
c.rhodes@imperial.ac.uk
Telephone: +44(0)2075941753
Fax: +44(0)2075940923

*Abstract* – The ubiquity of portable wireless-enabled computing and communications devices has stimulated the emergence of malicious codes (wireless worms) that are capable of spreading between spatially proximal devices. The potential exists for worms to be opportunistically transmitted between devices as they move around, so human mobility patterns will have an impact on epidemic spread. The scenario we address in this paper is proximity attacks from fleetingly in-contact wireless devices with short-range communication range, such as Bluetooth-enabled smart phones. An individual-based model of mobile devices is introduced and the effect of population characteristics and device behaviour on the outbreak dynamics is investigated. We show through extensive simulations that in the above scenario the resulting mass-action epidemic models remain applicable provided the contact rate is derived consistently from the underlying mobility model. The model gives useful analytical expressions against which more refined simulations of worm spread can be developed and tested.

# 1. Introduction

The modern world has become increasingly mobile. As a result, traditional ways of connecting computing devices to the Internet (and to each other) via physical cables have proved inadequate. Recent years have seen the widespread adoption of portable computing devices which are equipped with a short-range wireless technology such as WiFi [1] or Bluetooth [2]. Wireless connectivity is greatly advantageous as it poses no restriction on the user's mobility and allows a great deal of flexibility. At the same time the ability to wirelessly connect to the Internet and other devices, and to transfer data on the move, is opening opportunities for hackers to exploit such features for launching new and previously unexplored security attacks on computer and communication networks [3, 4, 5, 6].

Indeed, the last few years have seen the emergence of a new class of potentially destructive computer viruses that exploit such wireless capabilities in order to spread themselves between nearby devices, often without any active user involvement. One important feature of these new types of computer worms is that they do not require Internet connectivity for their propagation and therefore can spread without being detected by existing security systems. Another important feature is that, since they target portable devices, they can exploit the mobility of users for their spreading, in a way which shows interesting analogies with the spread of infectious diseases in a human population.

The spread of such wireless epidemics [4, 5] among WiFi-enabled computers placed at fixed locations has been investigated very recently [5]. These studies have revealed that the patterns of epidemic spreading in such networks are greatly different from the much studied epidemics in wired networks, and are strongly influenced by the spatial nature of these networks and the specifics of wireless communication.

The above static description is relevant in a situation where the underlying wireless contact network along which the epidemic spread is either connected or has a very large connected component. However, when device density is low or the infected devices have a very limited communication range (e.g. in the case of Bluetooth-enabled mobile phones [3]), at any time instant the underlying wireless contact network will be fragmented into many isolated clusters, rendering network propagation ineffective. In such situations, we expect the worm spreading to take place in an opportunistic manner where infected devices exploit the mobility of the users to transmit the worm to other devices to which they make a fleeting contact. The temporal patterns of such contacts and their duration depend on the underlying movement patterns of user mobility [7]. Therefore, in order to model the opportunistic spread of wireless worms between portable devices it will be necessary to develop a class of epidemic models that reflect the characteristics of human mobility, patterns of population aggregation and the wireless nature of worm transmission between devices [8]. Such models will inform the construction of mitigation strategies aimed at containing and eliminating future worm epidemics.

Here we present a model for the epidemic dynamics of a worm outbreak in a mobile spatially distributed population of wireless-enabled devices. Whilst worm spreading in fixed and ad-hoc networks has begun to be investigated in some detail [5, 6], the opportunistic spread between spatially proximal mobile devices has received less attention. The model elaborated below is constructed so as to reflect device mobility as well as the transmission characteristics of the worm. The transmission characteristics are determined by the wireless technology that is used for inter-device communication. Specifically, we show that in the above low density regime worm epidemics can be described by standard mass-action mixing models provided the contact rate is obtained self-consistently within the model itself.

The model framework presented here provides a number of analytic results, which are verified via individual-based simulations. These results are important in quantifying the impact of device mobility on the spreading of worms and other types of viruses in mobile wireless networks. Our results are also relevant to the analysis of novel delay-tolerant communication protocols which are being intensively researched for information dissemination and routing in intermittently connected wireless networks [9].

## 2. Opportunistic Transmission Model in a Mobile Population

### 2.1 Contact rate calculation

It is necessary to first calculate the contact rate between individuals in a mobile spatially distributed population. To do this, the motion-dependent contact rate for a given individual (denoted *i*) with others in the population is calculated [10, 11]. The contact process model consists of a population of individuals that are randomly and uniformly distributed over a two-dimensional domain with a density $\rho$. As illustrated in Figure 1 each individual moves independently of the others with a constant straight-line velocity $\vec{v}$, with their direction vectors distributed uniformly in azimuth in the plane. A specified individual, *i*, is introduced and moves through the domain with velocity $\vec{v}_i$. If one of the individuals passes within a radius $R$ of the specified individual *i* then, by definition, a contact has been made. Here we initially constrain $R \ll l$, where $l$ is the mean inter-particle spacing. Additionally we insist that the length scale of the domain $\Gamma \gg l$, giving $R \ll l \ll \Gamma$. (Later we investigate the effects of relaxing this constraint). Figure 2 shows the geometry of the interaction. In order for an individual to be exposed to contact by *i* during a time period $dt$ the individual must lie within the rectangular-shaped area swept by the motion of *i* lying along the direction of the vector $\vec{w} = \vec{v}_i - \vec{v}$. The area swept in this time period $dA = 2Rwdt$, where the relative speed is given by $w = \left(v_i^2 + v^2 - 2v_i v \cos\phi\right)^{1/2}$, where $\phi$ is the angle between the velocity vectors.

Given the independent motions of the particles, the density of individuals with velocity vectors with directions in the range $\phi$ and $\phi + d\phi$ is $\rho d\phi / 2\pi$ [12]. Therefore, the average number of individuals entering the swept area in $dt$ is $dN_\phi = \rho 2Rw d\phi dt / 2\pi$. In order to obtain the total number of individuals encountering *i* it is necessary to integrate over the planar angle $\phi$. Hence, the number of individuals entering the area around *i* bounded by the radius $R$, *i.e.* the mean contact rate *CR*, is given by:

$$CR = \int_0^{2\pi} \frac{dN_\phi}{dt} = \frac{\rho R}{\pi} \int_0^{2\pi} w \, d\phi \qquad (1)$$

Substituting for $w$ gives:

$$CR = \frac{\rho R}{\pi} \int_0^{2\pi} \left(v_i^2 + v^2 - 2v_i v \cos\phi\right)^{1/2} d\phi \qquad (2)$$

Equation 2 reduces to

$$CR = \frac{4\rho R}{\pi}(v_i + v) \int_0^{\pi/2} \left(1 - m\sin^2\omega\right)^{1/2} d\omega \qquad (3)$$

where $m = 4vv_i/(v+v_i)^2$. This is an elliptic integral of the second kind [13] and can be written in standard notation:

$$CR = \frac{4}{\pi} R\rho(v_i + v) E(m) \qquad (4)$$

To simplify what follow, assume that the index individual moves with the same speed $v$ as the rest of the population, so the contact rate can be written

$$CR = \frac{8}{\pi} R\rho v. \qquad (5)$$

It is generally the case that individuals in a population move with a range of different speeds. It has been shown in [11] that when the individual speeds are distributed according to a Maxwell-Boltzmann distribution, for example, equation 5 becomes

$$CR = \frac{8}{\pi} R\rho \bar{v} \qquad (6)$$

where $\bar{v}$ is the mean population speed, so it is possible to accommodate a distribution of individuals' speeds within this framework.

## 2.2 Worm transmission

If we now equate the distance *R* with the communication range of a wireless device and assume that there is a single infected device that is capable of transmitting a wireless worm with probability *p* to any other wireless device that finds itself within the range set by *R*, the rate of generation of new infected devices (*I*) is given by

$$\frac{dI}{dt} = \frac{8}{\pi} R \rho \bar{v} p \tag{7}$$

The secondary infectives go on to infect other uninfected devices, so when many infectives are present, then, assuming there is a density of $\rho_i$ infectives and $\rho_s$ uninfected devices (i.e. susceptibles *S*) in the domain, the rate of generation of infectives *per unit area* is given by:

$$\frac{1}{A}\frac{dI}{dt} = \frac{8R\bar{v}p}{\pi} \rho_s \rho_i \tag{8}$$

It is often easiest to consider changes to the population sizes when investigating the dynamics of the infection, so converting equation 13 to population sizes results in:

$$\frac{1}{A}\frac{dI}{dt} = \frac{8R\bar{v}p}{\pi} \frac{S}{A}\frac{I}{A} \tag{9}$$

And this reduces to

$$\frac{dI}{dt} = \frac{8R\bar{v}\rho p}{\pi} \frac{SI}{N} \tag{10}$$

or

$$\frac{dI}{dt} = \beta \frac{SI}{N} \quad (11)$$

where

$$\beta = \frac{8R\bar{v}\rho p}{\pi} \quad (12)$$

Equation 11 is the standard frequency dependent mass-action assumption [14] that is widely used in epidemic modelling, where $\rho$ is the total population density and $N$ is the total population size, i.e. $\rho = N/A$. However, in epidemic models, the transmission rate, $\beta$, is usually determined by the empirically observed "basic reproductive rate" of the pathogen, whereas here we are able to relate this term to the underlying population motion and transmission behaviour.

**2.3 Refinement of the basic model**

In the basic contact process described above, any uninfected wireless device that passes within the communication radius $R$ of an infected device for a fleeting duration is just as likely to get infected (with probability $p$) as one that passes very close to the infective. In practice, however, it is likely that the longer an uninfected device is within range of an infected one the greater the probability that a worm transmission event will occur. To address this refinement, we introduce the idea of transmission profile.

The transmission profile will be determined by the radial decay of radio frequency signal strength within the region bounded by $R$. Any susceptible traversing a straight-line path within $R$ will experience an exposure to worm transmission that is both dependent upon this profile and its path length through the domain bounded by $R$.

Susceptibles passing closer to an infective will spend longer close to an infective *and* be exposed to higher integrated signal strength than a susceptible further away. Figure 3 shows two individuals traversing the communication region of a wireless device but with differing closest points of approach.

Denoting the path length function by $l(r)$, and the transmission profile by $f(r)$, the probability that an uninfected device gets infected $p(r) \propto l(r)f(r)$. For a circular domain of radius $R$, $l(r) = 2(R^2 - r^2)^{1/2}$.

Usually, the transmission profile will be proportional to the RF signal power at the receiving device, which is typically modelled as $\propto P_{trans}/cr^\alpha$ where $c$ and $\alpha$ are environmentally dependent [15]. For simplicity, adopting the transmission model described by Glauche *et al.* [16] and Nekovee [5], we assume that each wireless device can establish links with devices within the transmission range $R$, whereas for devices beyond $R$ connection is not possible. In this case the transmission profile is independent of the radial distance within $R$ so $f(r) = k$. Consequently, the probability that an uninfected device becomes infected is $p(r) = 2\left(R^2 - r^2\right)^{1/2} k$. To normalise the probability distribution, if an uninfected device collides with an infective there is a probability $p$ that the susceptible becomes infected, i.e. $p(0) = p = 2Rk$, so the constant $k = p/2R$. Therefore, $p(r) = \frac{p}{R}\left(R^2 - r^2\right)^{1/2}$. The overall transmission rate for the uniform contagion profile is

$$\beta = \frac{8}{\pi}\rho\bar{v}\int_0^R \frac{p}{R}\left(R^2 - r^2\right)^{1/2} dr \qquad (13)$$

which gives

$$\beta = 2R\rho\bar{v}p \qquad (14)$$

Therefore, taking into account the time spent by a susceptible within $R$ reduces the rate of production of infectives to around 80% of that seen in the basic contact process model (c.f. equation 12). It is possible that more complex radial dependencies for the transmission model might be applicable depending upon the environment in which the transmissions are being made. The adoption here of the transmission model presented by Nekovee [5] permits an analytic expression for the transmission parameter to be derived. More complex transmission models can be straightforwardly accommodated within this framework, but a numerical evaluation of the necessary integrals would be required.

## 3. Implications for the epidemiology of wireless worms

Above we have introduced a model for the spread of malicious code entities between wireless enabled devices. The model has been constructed with the explicit recognition of device mobility. Using the model it is possible to explore the behaviour of worm epidemics in mobile devices.

Consider a population of $N$ individual devices existing at a density $\rho$ and moving with a mean speed $\bar{v}$. Assume each device has a wireless communications radius $R$. If a worm is introduced into the system each device can be either uninfected (susceptible, $S$), infected ($I$) or recovered ($P$) (i.e. a "patch" has been added to render the device unable to infect other devices and no longer capable of being re-infected).

In a finite population of fixed size, the worm epidemic is described by

$$\frac{dS}{dt} = -2R\rho\bar{v}p\frac{SI}{N} \tag{15}$$

$$\frac{dI}{dt} = 2R\rho\bar{v}p\frac{SI}{N} - \delta I \tag{16}$$

$$\frac{dP}{dt} = \delta I \tag{17}$$

For an epidemic to take off, the initial infected device needs to generate a positive number of infected devices, i.e. $dI/dt > 0$ at $t \approx 0$. For this condition to hold

$$2R\rho\bar{v}p > \delta \qquad (18)$$

In this scenario the epidemic will always die out eventually ($I \to 0, t \to \infty$), but the sum of those devices in the population that get infected with the worm is given by the solution of the transcendental equation

$$P_\infty = N\left(1 - e^{-\frac{2R\rho\bar{v}p}{\delta N}P_\infty}\right) \qquad (19)$$

Where $P_\infty$ is the limiting number of recovered devices as $t \to \infty$ (i.e. the sum of all those infected during the course of the epidemic). Note that there is a critical device density $\rho_c = \delta/2R\bar{v}p$ below which no worm epidemic is possible.

From this it is apparent how the proportion of devices infected during an outbreak is dependent on the model parameters. Figure 4 shows a comparison of the time-dependence of the number of infectives from the model (equations 15-17) with a simulation of the process. In this simulation we took a typical urban population density of 3000 individuals/km$^2$ with a mean speed of around 2km/day, with $p$=0.1. It was assumed that each wireless device has a transmission radius of 5m, and such a transmission range corresponds to that seen in a Class 2 Bluetooth device [2, 7]. In this case the epidemic dynamics result from the aggregated effect of numerous dyadic transmission events, because at any given time each device will almost never have more than one other wireless device within its transmission range $(\pi 4 \times 3 \times 10^{-3} \approx 0.2)$. Channel access protocols such as medium access control (MAC) [5] have minimal impact, as there are rarely any other competing devices within the transmission range of an infected device. Consequently, the epidemic characteristics are well described by a standard mass-action model.

Next, it is of interest to investigate what happens as the wireless device transmission range $R$ is increased from 10m to 40m, which is more typical of Class 1 Bluetooth devices. Figure 5 shows a series of epidemic curves for increasing wireless transmission range, $R$, and using the same simulation parameters as previously. As $R$ increases the epidemic dynamics, although occurring on a faster time-scale, remain mass-action like. The results of a basic mass-action model (equations 15-17) are shown for comparison and the agreement between simulation and theory is good. Although each simulation is a single realisation of what is a stochastic process, at this level the fluctuations are minimal. As $R$ gets large the beginnings of a deviation between the SIR model and the simulation is becoming evident for this single realisation of the dynamics.

In addition to the radius of wireless transmission, $R$, there is also the probability of worm transmission, $p$. It is possible that a channel allocation protocol such as MAC might serve to lower the transmission probability, $p$, as $R$ increases. This is because increasing numbers of devices find themselves within the transmission range, so the listen-before-talk aspect of MAC reduces the opportunities for transmission. In effect, as $R$ increases $p$ decreases, so (depending on the specific details of the protocol) the product $Rp$ in equation 14 may remain relatively constant. Therefore, as the transmitter range increases, if a channel allocation protocol is in operation, there may not be the expected increase in the transmission rate of the pathogen. This effect is rather like the "self-throttling" behaviour noted by Nekovee [5] in the spread of worms in wireless ad-hoc networks. Channel allocation protocols serve to inhibit worm spread and could be regarded as a potential means to intervene in eliminating worm epidemics.

From this it is possible to see that within the same population, the resultant worm dynamics will be dependent upon the characteristics of the devices. Widely different epidemic outbreaks are possible as the parameters relating to the devices changes. In real populations the transmission range of each device may well be influenced by the local environment (i.e. by the presence of wall, buildings etc) so it is possible that not all identical wireless devices will have a similar communication properties.

**Conclusions**

The threat of attack from malicious code is rapidly increasing as wireless technology becomes routinely packaged in information processing and communication devices. Understanding the dynamics of these wireless worm epidemics will be essential to devising efficient intervention schemes to protect users and maintain network functionality. Just as in human disease epidemiology, mathematical modelling has a key role to play in developing this understanding of worm epidemiology. Using an individual-based approach we have developed a contact process model for the interaction of a population of interacting mobile devices. The model shows how the epidemic dynamics are influenced by factors such as host speed, worm transmission probability and device transmission range. Specifically, we have shown that both for short-range and longer range transmitters the epidemic dynamics in the low density limit are captured by a standard mass-action epidemic model where the transmission rate is set by the factors such as population density, device transmission range and speed. Agreement between the simulation of worm spread in a mobile population and a simple epidemic model is good. However, as device transmission range increases the impact of channel allocation algorithms plays an increasing role in influencing the probability of worm transmission (this is particularly the case for WiFi communication, where devices have to contend for a limited number of communication channels). Such protocols serve to inhibit the rate of epidemic spread and serve to reduce the overall transmission term for the epidemic.

Wireless epidemics will reflect the aggregation and movement of humans, so many of the techniques of conventional epidemiology will be needed to better reflect outbreaks. In the model here, future refinements are needed to better understand the impact of more complex movement patterns (other than straight-line) and to consider inhomogeneities in the distribution of the population. Also, in parallel with these developments, the effect of channel allocation protocols needs to be investigated in more detail. Furthermore, when the density of devices is sufficiently high such that large network clusters can emerge, the spreading dynamics becomes increasingly more affected by network correlation effects, which are not included in the standard mass action epidemic models. Such effects, however, can be described using models from spatial epidemics, and will be considered in future work.

**Figure Captions**

Figure 1:    A specified individual is located at the centre of the large (dashed) circle, radius $R$, and is moving through the domain in a straight line. Other individuals move across the domain in straight lines; velocity vectors are uniformly distributed in azimuth. When an individual passes within a distance $R$ of the specified individual contact is made with probability $p$. All the particles are treated as point-like.

Figure 2:    Geometry of the interaction between a specified individual and another individual in the population. The movement of all the individuals takes place in the $x$, $y$ plane. A specified individual, $i$, situated at the co-ordinate origin moves with velocity vector $\overline{v_i}$ (dashed line) along the negative $y$ axis, whilst another individual with which it will make contact moves with velocity vector $\overline{v}$. These vectors are at an angle $\varphi$. The vector $\overline{w}$ is the relative velocity vector of the two individuals. The area of hazard for the non-specified individual ($dA = 2Rwdt$) is given by the lozenge-shaped area minus the semicircular end pieces.

Figure 3: Two devices passing within $R$ but on different trajectories. They are exposed to differing risk of infection due to the differing time they spend within the region $R$ centred on an infected wireless device.

Figure 4: Infection curve for the epidemic model and a simulation ($R = 5$m). The time dependence of the number of infected devices from the simulation is shown (dots). The mass-action model (solid line) is shown for comparison.

Figure 5a-c: Infection curves for $R = 10$m, 20m, and 40m. Simulated results (dots) and mass action model (solid line) are shown

Figure 1:

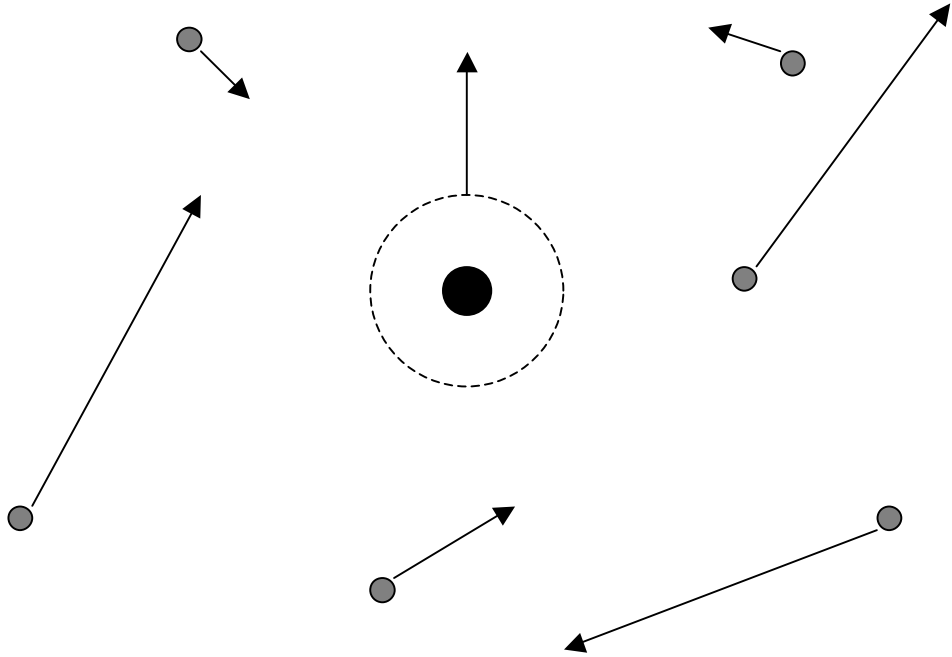

Figure 2:

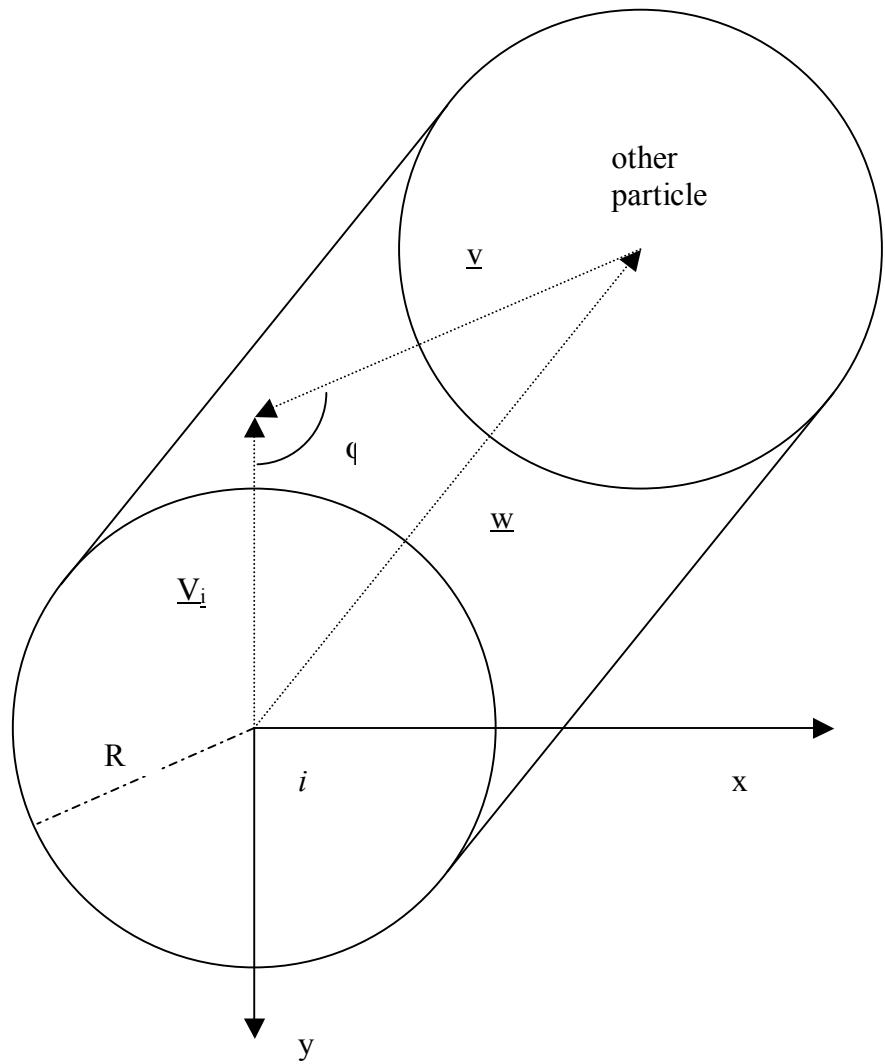

Figure 3:

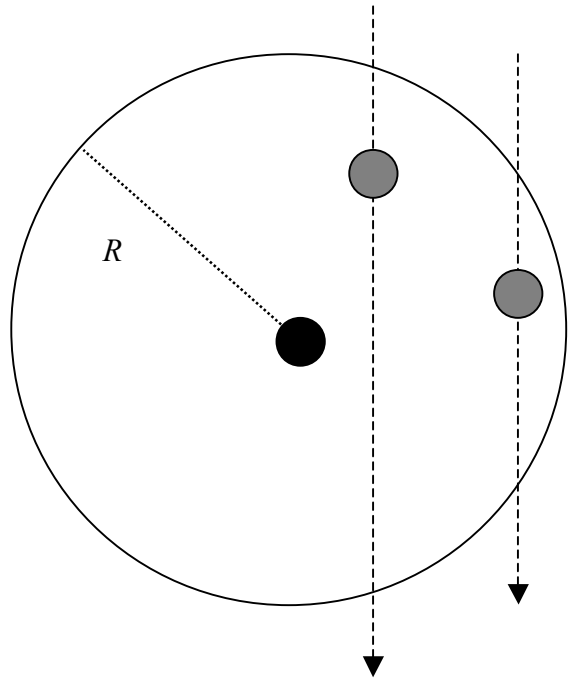

Figure 4:

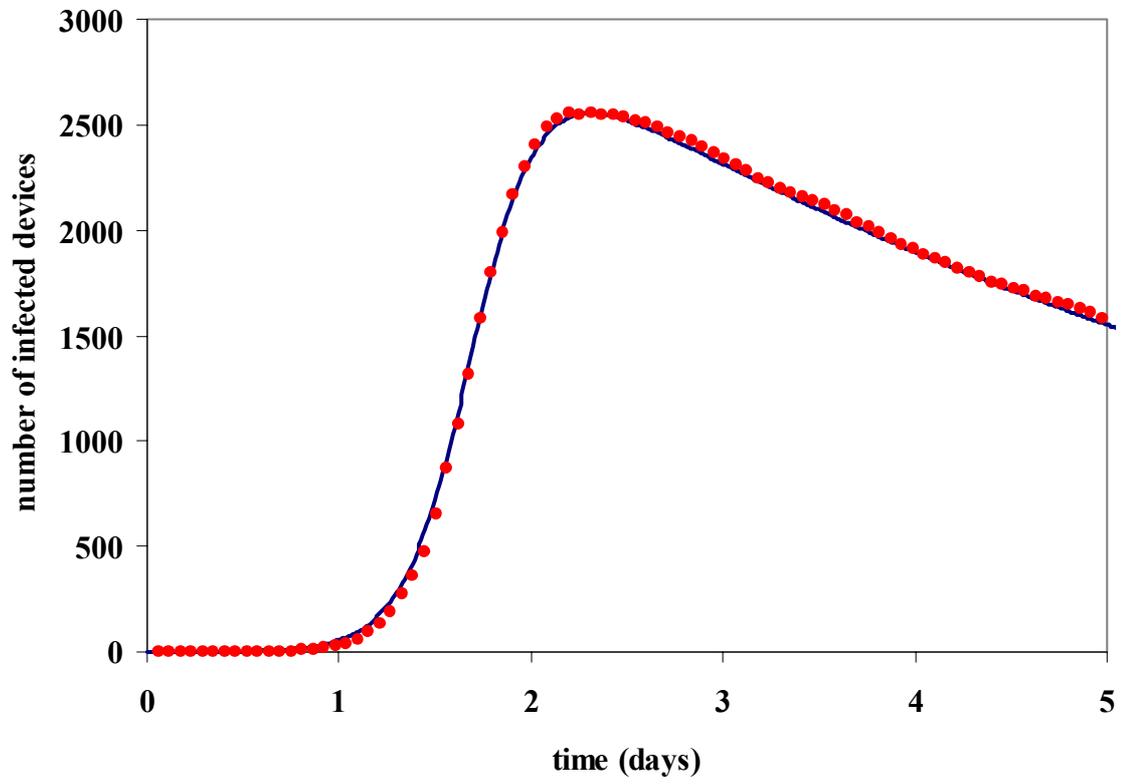

Figure 5a

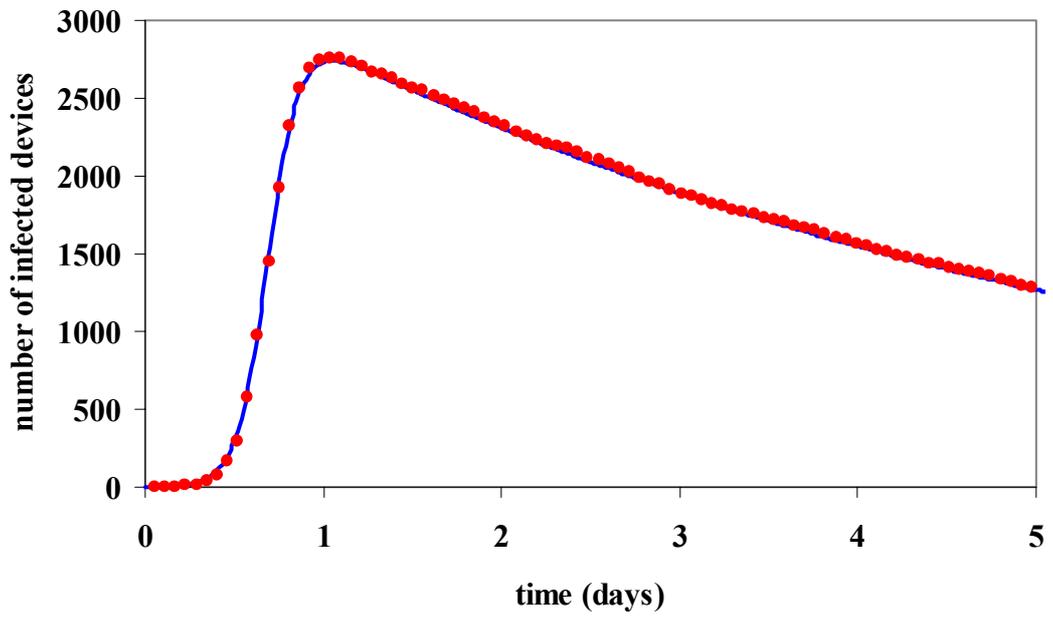

Figure 5b

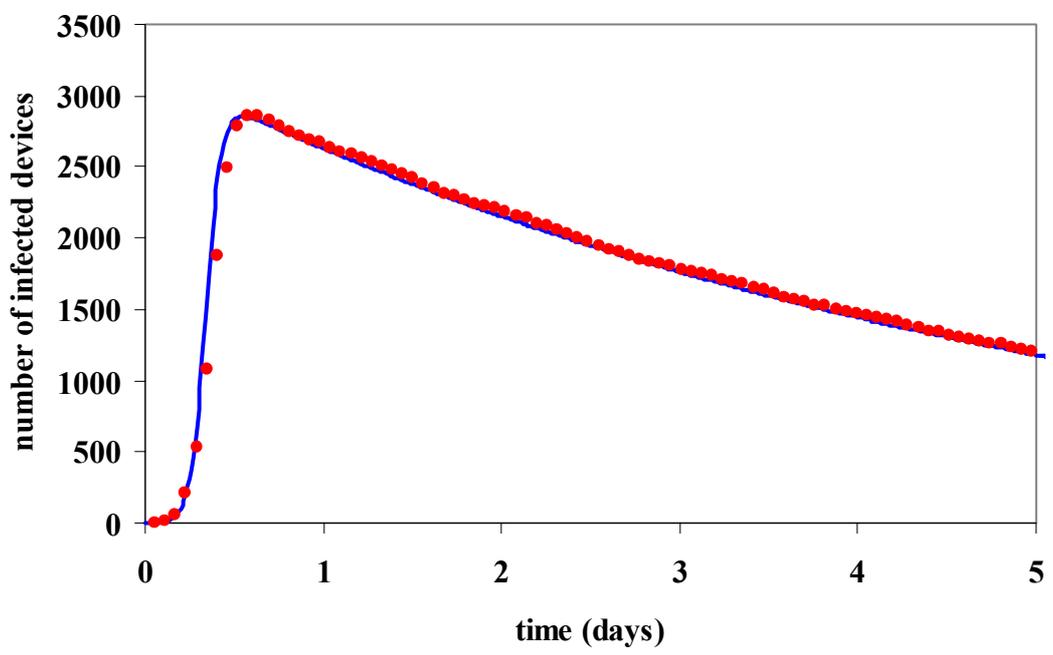

Figure 5c

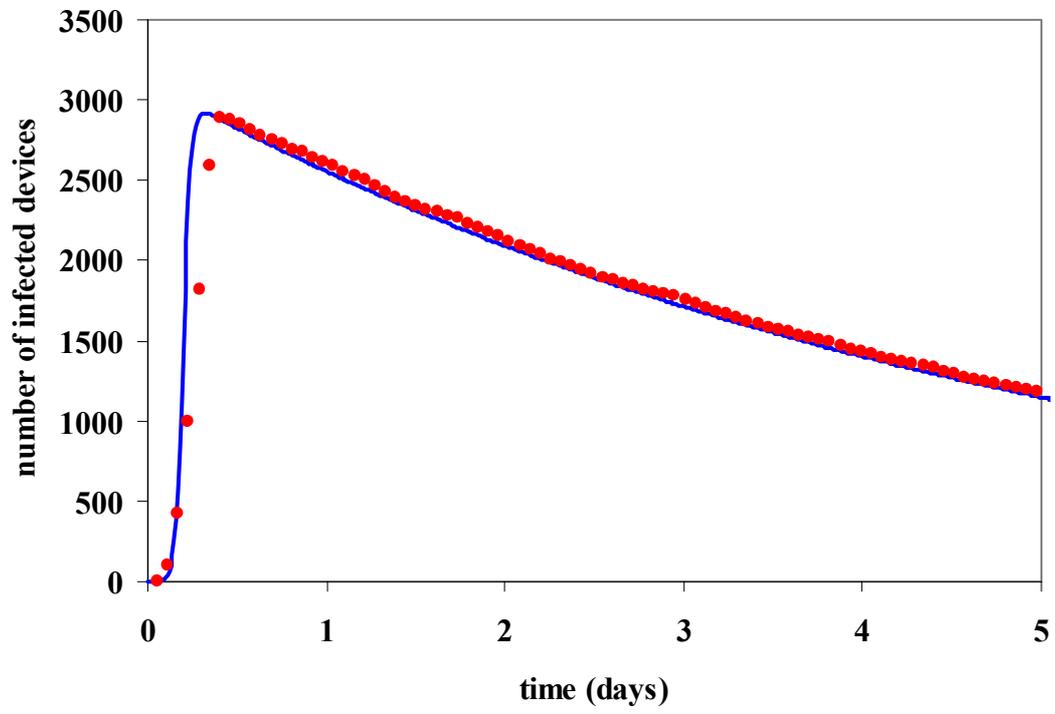